# Hydrogen storage in pillared Li-dispersed boron carbide nanotubes

*Xiaojun Wu, Yi Gao and Xiao Cheng Zeng*[*]

Department of Chemistry and Nebraska Center for Materials and Nanoscience, University of Nebraska-Lincoln, Lincoln, Nebraska 68588

**Abstract**

Ab initio density-functional theory study suggests that pillared Li-dispersed boron carbide nanotubes is capable of storing hydrogen with a mass density higher than 6.0 weight% and a volumetric density higher than 45 g/L. The boron substitution in carbon nanotube greatly enhances the binding energy of Li atom to the nanotube, and this binding energy (~ 2.7 eV) is greater than the cohesive energy of lithium metal (~1.7 eV), preventing lithium from aggregation (or segregation) at high lithium doping concentration. The adsorption energy of hydrogen on the Li-dispersed boron carbide nanotube is in the range of 10 – 24 kJ/mol, suitable for reversible $H_2$ adsorption/desorption at room temperature and near ambient pressure.

Hydrogen is a promising alternative fuel to reducing our dependence on fossil fuels [1-2]. The implementation of hydrogen fuels in transportation application however is currently limited by the lack of an economic method for hydrogen storage. The identification of materials that can store hydrogen at both high gravimetric and volumetric density near ambient conditions represents a critical challenge to the commercialization of fuel-cell powered automobiles. On-vehicle hydrogen storage not only requires operation within minimum volume and weight specifications but also requires fast-recycling of hydrogen near ambient temperature and pressure. Adsorptive hydrogen storage has been perceived as a viable option to meet Department of Energy (DOE)'s 2010 target, that is, 6.0 weight% in gravimetric capacity and 45 g/L in volumetric density. Among other adsorptive media, microporous metal-organic frameworks and carbon-based nanostructures are most attractive for hydrogen storage. Major progresses have been made recently in designing metal-organic frameworks that can achieve high gravimetric and volumetric density of hydrogen at 77 K [3-6]. Carbon based materials such as carbon nanotubes are also capable of high uptake of hydrogen because of their high surface area and light weight [7-13]. However, at room temperature and ambient pressure the hydrogen-storage capability of pristine carbon nanotubes has been shown to be too small to meet the DOE's target [14].

Previous experimental and theoretical studies have also shown that doping alkali or transition metal atoms on carbon nanotube [14-22] can appreciably increase hydrogen uptake largely due to the enhanced adsorption energy of $H_2$ with the metal-doped carbon nanotube. Yildirim *et al.* studied the interaction between hydrogen molecules and Ti-coated single-walled carbon nanotubes (SWCNTs) [18]. They found that $H_2$ can be adsorbed on the Ti atom with enhanced adsorption energy. Up to 8 wt% of hydrogen can be taken up if the Ti atoms can be uniformly coated on the SWCNT surface. Lee *et al.* explored hydrogen adsorption on Ni-dispersed SWCNTs and estimated that the SWCNTs can release 10 wt% of hydrogen at room temperature [22]. Further studies of hydrogen adsorption on Pt-doped SWCNTs [19] and organometallic fullerenes [24] were reported by several groups, and high hydrogen uptakes were predicted theoretically. In all studies, transition metal atoms were assumed to be uniformly coated on the SWCNT surface or fullerenes. As such, every metal atom can adsorb several hydrogen molecules. Recently, Sun *et al.* investigated interaction between fullerene $C_{60}$ and transition metal Ti [25]. They showed that twelve Ti atoms tend to form a small cluster rather than uniformly dispersed on the $C_{60}$ surface. They further studied the Li-dispersed $C_{60}$ and found that twelve Li atoms tend to locate on the twelve pentagons $C_{60}$ rather than forming a cluster, due to slightly higher binding energy between Li and $C_{60}$ than the cohesive energy of lithium metal [26]. Up to 120 hydrogen atoms may be physisorbed on a $Li_{12}C_{60}$ cluster. A main challenge in mass production of the fullerene-based hydrogen-storage media is to avoid aggregation of the $Li_{12}C_{60}$ clusters [26].

On alkali metal-doped carbon-nanotube, an early study reported large hydrogen-storage capacities at room temperature [14]. Later experiments indicated that the high hydrogen uptakes were likely due to water impurities presented in the system [15,16]. Only up to 2 wt% hydrogen uptake has been achieved at room temperatures. Several theoretical studies have predicted that the adsorption energy of $H_2$ on Li-dispersed carbon nanotube can be substantially enhanced [17, 20, 21]. For example, Deng *et al.* have shown that Li-dispersed pillared SWCNTs with Li:C ratio 1:3 can give rise to high hydrogen uptake (6.0 wt%) at room temperature and 50 bars, assuming a portion of $H_2$ can be stored inside the tubes [17]. To achieve higher lithium doping concentration without lithium segregation, the binding energy between the lithium and nanotube should be further enhanced to exceed the cohesive energy of bulk lithium. Herein, we propose to use pillared Li-dispersed boron carbide nanotubes to achieve high lithium coverage as well as high hydrogen uptake. It is known that substitution of boron atoms in carbon nanotube can modify the electronic properties and chemical reactivity of the carbon nanotube [28]. Peng *et al.* found that B substitution in the carbon nanotube can enhance the adsorption energy of CO and water [28]. Kim *et al.* also demonstrated enhanced adsorption energy of $H_2$ on boron-doped fullerence [29]. Zhou *et al.* showed theoretically that the Li binding energy can be enhanced through B doping in SWCNTs [30]. We note that several types of boron carbide (BC) nanotubes have been fabricated in the laboratory such as $BC_3$ and $(B_xC, x \leq 0.1)$ nanotubes [31-38]. Thus, it is possible to design and produce pillared boron carbide nanotubes.

**Theoretical Calculations**

We performed all-electron density-functional theory (DFT) calculations with double numerical basis sets and polarization function (DNP basis set) implemented in the DMol3 package [39-41]. Spin-unrestricted DFT in the generalized-gradient approximation (GGA) with the Perdew-Burke-Ernzerhof (PBE) functional [42] as well as in the local density approximation (LDA) with the Perdew-Wang (PWC) functional were adopted [43]. Because the interaction between $H_2$ and Li-dispersed carbon nanotube is physisorption in nature, we employed both GGA and LDA as a way to estimate the lower/upper bound for the physisorption energy. We note that a previous quantum Monte Carlo calculation shows that the adsorption energy of hydrogen in carbon materials is typically in between the calculated adsorption energy based on GGA and LDA [24]. Note also that another previous *ab initio* study using the Møller-Plesset second-order perturbation (MP2) method [23] shows that GGA tends to underestimate the adsorption energy of $H_2$ on SWCNT whereas LDA gives reasonable adsorption energy compared to the MP2 calculation [20, 23, 30].

We chose a single-walled zigzag (8,0) carbon nanotube as the model prototype. A tetragonal supercell of 30 Å × 30 Å × 8.54 Å was used to simulate the nanotube with infinite length, where the supercell length in the axial direction 8.54 Å is about twice of the periodic length of the zigzag (8,0) carbon nanotube. The single-walled (8,0) BC nanotube was constructed via the boron substitution of the corresponding carbon nanotube. The periodic length along the BC tube axis is 4.78 Å. In the periodic DFT calculation, the real-space global cutoff radius was set to be 5.1 Å. The Brillouin zone was sampled with a *k*-points separation of 0.04 (Å$^{-1}$) using the Monkhorst-Pack scheme [44]. Test calculations reveal that increasing *k*-points does not affect the results.

**Results and Discussion**

We first evaluated the binding energy of a Li atom (per supercell) to the (8,0) carbon nanotube. The Li atom favors to locate at the hollow site of the hexagonal ring on the outer surface. Two types of Li-C bonds exist with bond length 2.185 and 2.383 Å (PBE calculation), respectively. The binding (or adsorption) energy of the Li atom is defined as $E_{ads}= E_{total}$[carbon nanotube] + $E_{total}$[Li] – $E_{total}$[carbon nanotube +Li], where $E_{total}$ is the total energy of the system per supercell. The calculated binding energy of the Li atom to the carbon nanotube is 1.74 eV (PBE calculation). The doping of Li atom modifies the electronic structures of the carbon nanotube that the Fermi level is shifted upward and crosses the original conduction band, resulting in a semiconductor-to-metal transition. Moreover, the contribution from the Li atom to the density of states is mainly in the region of 2.0 eV above the Fermi level. The Hirshfeld charge analysis indicates that about 0.45 $e$ electrons are transferred from the Li atom to carbon nanotube, resulting in partially cationic Li atom. The transferred electrons to the carbon nanotube elevate the Fermi level and lead to the semiconductor-to-metal transition.

Rao and Jena showed that a Li cation can adsorb at least six hydrogen molecules [45]. The strong molecular adsorption stems mainly from the electrostatic interaction, that is, the polarization of hydrogen molecules by the Li cation. On carbon nanotube's surface the Li atom is partially ionized. Thus, the Li atom can adsorb up to four $H_2$ as shown in Fig. 1. The average adsorption energies per $H_2$, based on PBE/GGA and PWC/LDA calculations, are summarized in Table I. As expected, the LDA gives slightly stronger interaction between $H_2$ and Li-doped carbon nanotube than the GGA. For example, on one $H_2$ adsorption, the adsorbed molecule has a H-H bond length of 0.757 Å [Fig. 1(a)], slightly larger than 0.752 Å of a free $H_2$. The corresponding adsorption energy (GGA) is 0.162 eV. The Hirshfeld charge analysis indicates that 0.079 $e$ electrons are transferred from $H_2$ to Li. The Li atom still carries a positive charge of 0.360 $e$. In contrast, the LDA calculation shows that the H-H bond is elongated from 0.766 Å (free $H_2$) to 0.775 Å upon adsorption, with the adsorption energy of 0.256 eV (Table I). Moreover, the Li-H distances calculated from the LDA are shorter than those from the GGA.

Both LDA and GGA predict that up to four $H_2$ molecules can be adsorbed on the Li atom, and the average adsorption energy per $H_2$ decreases with increasing the number of $H_2$.

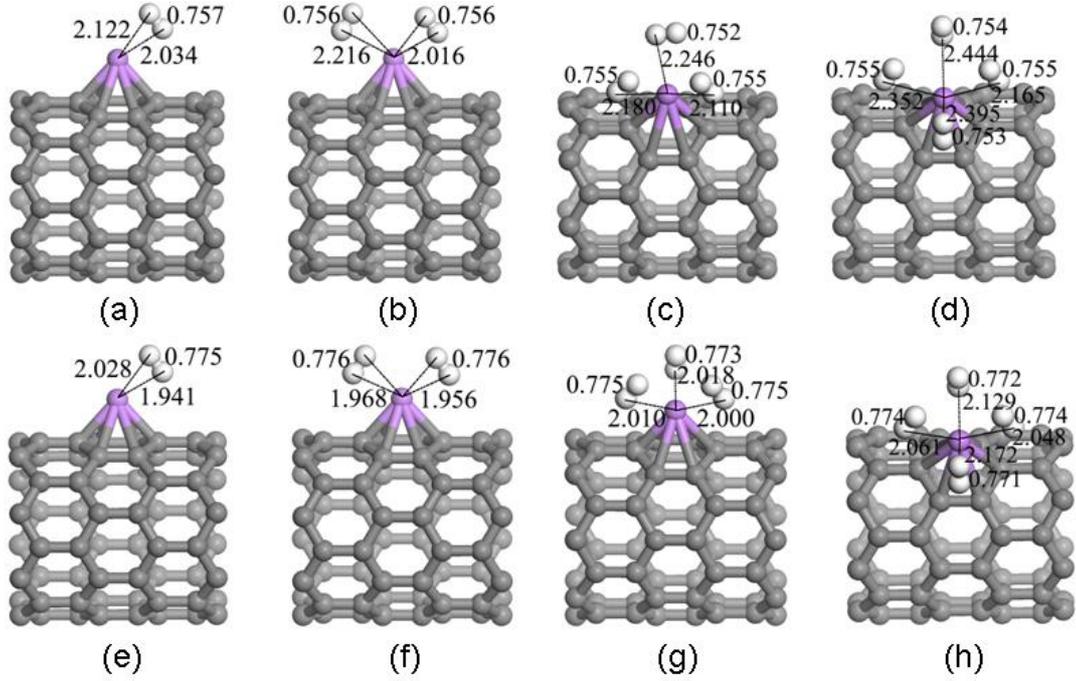

**FIG. 1. Optimized structures of Li-doped (8, 0) SWCNT with one to four hydrogen molecules adsorbed. (a) to (d) are optimized structures based on PBE/GGA calculation, and (e) to (h) based on PWC/LDA calculation. All bond lengths are in Å.**

**Table I. Average adsorption energies per $H_2$ on Li-doped SWCNT (Li-SWCNT) and Li-doped SWCNT with one B atom substitution (Li-CNT-B), calculated based on PBE/GGA and PWC/LDA.**

|  | PBE/GGA (eV) | PWC/LDA (eV) |
|---|---|---|
| 1 $H_2$@Li-SWCNT | 0.162 | 0.256 |
| 2 $H_2$@Li-SWCNT | 0.143 | 0.259 |
| 3 $H_2$@Li-SWCNT | 0.121 | 0.242 |
| 4 $H_2$@Li-SWCNT | 0.097 | 0.213 |
| 1 H$_2$@Li-CNT-B | 0.147 | 0.247 |
| 2 H$_2$@Li-CNT-B | 0.126 | 0.234 |
| 3 H$_2$@Li-CNT-B | 0.106 | 0.218 |
| 4 H$_2$@Li-CNT-B | 0.084 | 0.196 |

Bhatia and Myers recently showed that the optimal adsorption energy for adsorptive hydrogen storage is ~15 kJ/mol or 0.15 eV per H$_2$ [27]. This value of adsorption energy was predicted to maximize the amount of adsorbed hydrogen molecules accessible at room temperature. As shown in Table I, the predicted adsorption energy of H$_2$ on the Li-doped SWCNT ranges from 0.1 – 0.26 eV, and their mean value is close to 0.15 eV. Indeed, Li-dispersed pillared SWCNTs have been predicted to be a promising hydrogen storage medium [26]. We also calculated the cohesive energy of bulk lithium, which is 1.70 eV (PBE). This cohesive energy is only slightly less than the binding energy of Li atom to the SWCNT, 1.74 eV (PBE). Hence, there is a possibility that as the lithium doping concentration increases, some Li atoms may aggregate into clusters. The aggregation will effectively reduce the coverage of lithium on the pillared SWCNTs. Deng *et al.* showed that the optimal lithium doping concentration is Li:C=1:3 in carbon-based materials [26]. Higher lithium doping concentration may lead to lithium segregation. To further increase the lithium doping concentration, we propose to consider boron substitution in SWCNT. It is known that boron substitution can significantly enhance the binding energy of Li atom with carbon nanotube [28,30]. In a test calculation, we replaced one carbon atom by a boron atom in the supercell (BC$_{63}$). The Li atom is energetically more favorable to be chemisorbed at the hollow site of C$_5$B hexagon ring. The calculated binding energy of Li atom is 2.727 eV, much larger than the cohesive energy of bulk lithium 1.70 eV. We also calculated the binding energy of a Li atom to the armchair (5, 5) tube and found that the binding energy is changed from 1.537 eV to the carbon nanotube to 2.397 eV to the boron-doped carbon nanotube. Lastly, we considered boron substitution of two separated carbon atoms (B$_2$C$_{62}$) [46]. The calculated binding energy of Li atom is about 2.647 eV, still much larger than that (1.74 eV) to the pristine carbon nanotube.

Considering one boron atom substitution (per supercell), we calculated hydrogen adsorption energies on Li-doped nanobute (see Table I). The average adsorption energy of $H_2$ ranges from 0.08 eV to 0.25 eV, depending on the number of adsorbed hydrogen molecules and the DFT method (GGA or LGA). Considering four hydrogen molecules adsorbed on a single Li atom, Hirshfeld charge analysis shows that the Li atom carries a positive charge of 0.43 $e$, slightly less than that on the pristine carbon nanotube. The charge on the B atom changes from +0.038 $e$ to -0.0008 $e$, due to the doping of Li atom. The results of these test calculations are encouraging as they suggest that the much enhanced binding energy between the Li atom and boron-substituted carbon nanotube can potentially elude the lithium aggregation problem at high lithium doping concentration, and yet still give a mean hydrogen adsorption energy close to 0.15 eV.

Next, to study hydrogen adsorption on pillared Li-dispersed boron carbide (BC) nanotubes, we constructed an ideal (8,0) BC nanotube, where each hexagon ring contains three boron atoms. Geometric relaxation using the PBE/GGA method shows that the (8,0) BC nanotube has a periodic length of 4.780 Å in the axial direction. The BC nanotube contains two types of B-C bond with the bond length of 1.560 and 1.509 Å, respectively. About 0.08 $e$ electrons are transferred from B to C atom, and the electronic structure calculation suggests that the (8,0) BC tube is metallic. An ideal Li-dispersed BC tube system is shown in Fig. 2, where the Li atoms occupy all the hollow sites of hexagon rings. After structure optimization, the Li-B bond lengths range from 2.104 to 2.208 Å, while the Li-C bond lengths range from 2.016 to 2.138 Å. The average binding energy of the Li atom to the BC nanotube is 3.10 eV, significantly larger than the cohesive energy of bulk lithium. Every Li atom carries a positive charge ~0.22 $e$. We then studied a perfect hydrogen adsorption cconfiguration such that every Li atom adsorbs one $H_2$ molecule. The entire chemical system can be described as $BCLiH_2$. The optimized structure of the $BCLiH_2$ system is shown in Fig. 2(b), where the Li-H distance ranges from 2.015 to 2.082 Å. The average hydrogen adsorption energy is 0.106 eV (GGA) or 0.243 eV (LDA). The realistic hydrogen adsorption energy is likely in between the GGA and LDA values.

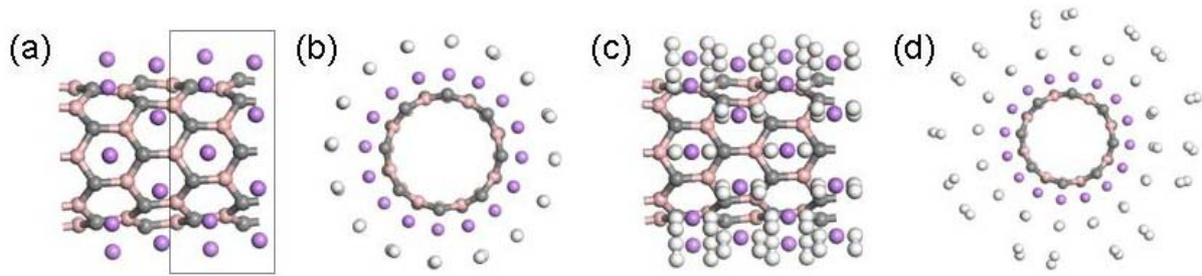

**FIG. 2.** (a) A side view of the optimized structure of Li-dispersed BC (8,0) nanotube (BCLi system). The rectangle frame highlights the unit cell of the system. (b) A top view of Li-dispersed BC nanotube with one $H_2$ adsorbed on every Li atom. (c) A side view and (d) a top view of an initial structure of the Li-dispersed BC nanotube with two $H_2$ adsorbed on every Li atom.

Can more than one hydrogen molecule be adsorbed on every Li atom for the Li-dispersed BC nanotube? To address this question, we considered an adsorption configuration such that every Li atom adsorbs two $H_2$ molecules. Initially, all $H_2$ molecules were placed on the Li atom with a Li-H distance less than 1.0 Å (see Fig. 2(c)). After full structural optimization, about half of total $H_2$ molecules are still adsorbed on the Li atoms with Li-H distance ranging from 2.021 to 2.110 Å, similar to the relaxed $BCLiH_2$ system. However, the other half are desorbed from Li atoms, as shown in Fig. 2(d). These desorbed hydrogen molecules, having Li-H distances longer than 4.6 Å, form a molecular layer that covers the inner $BCLiH_2$ system. The adsorption energy of the outer $H_2$ is about 0.015 eV per $H_2$ (GGA), stemming mainly from weak van der Waals interaction. Thus, although a Li atom doped on the BC tube is capable of adsorbing four $H_2$ molecules, Li-dispersed BC nanotube can only adsorb one $H_2$ per Li atom. This limitation of $H_2$ uptake is due to *steric repulsion* between congested $H_2$ molecules because of the short Li-Li distance (3.02 Å) at the full lithium coverage. Still, with only one $H_2$ uptake per Li atom, the $BCLiH_2$ system has a gravimetric density of hydrogen 6.34 wt%, exceeding the 2010 DOE target.

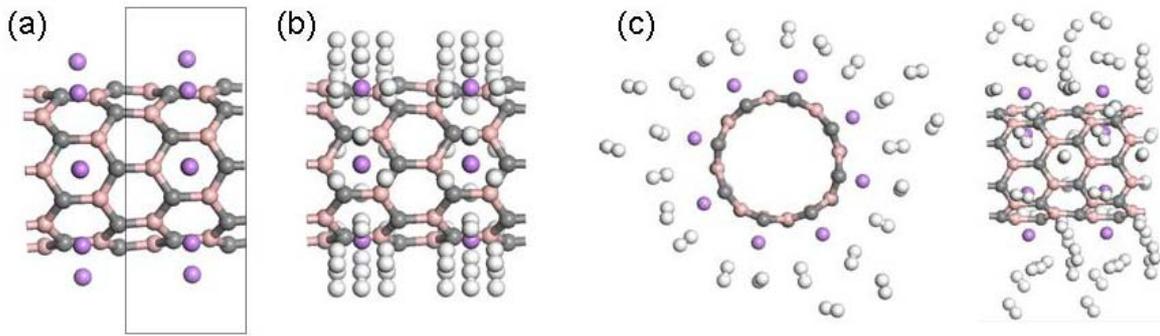

**FIG. 3.** A side view of (a) the optimized structure of a BC nanotube with every other hexagon layer occupied by Li atoms ($B_2C_2Li$ system). (b) The initial configuration with three $H_2$ molecules placed on each Li atom. (c) The top and side view of the optimized structure based on the initial configuration (b).

We also studied a model system $B_2C_2Li$, as shown in Fig. 3(a), which entails half of the lithium doping ratio (Li:C) than the BCLi system. In this model system, every other hexagon layer is covered by Li atoms. Although the number of Li atoms is reduced by half, the spatial volume around every Li atom is effectively larger. In this case, the binding energy of the Li atom is 3.242 eV (GGA). We then placed three $H_2$ on each Li atom. The initial adsorption configuration with total 24 $H_2$ per supercell is shown in Fig. 3(a) where the supercell is highlighted by a rectangle. After structural optimization, three locations for $H_2$ were observed: (1) $H_2$ adsorbed on the Li atom with Li-H bond length ranging from 2.062 to 2.191 Å, (2) $H_2$ adsorbed in between two nearest-neighbor Li atoms with Li-H bond length ranging from 3.021 to 3.632 Å, and (3) desorbed $H_2$ with Li-H distance larger than 4.2 Å. Interestingly, after we removed all the desorbed hydrogen molecules from the system, we found that the gravimetric density of the hydrogen is still 7.94 wt%, and the average hydrogen adsorption energy is 0.073 eV (GGA). This model study suggests that even if the BC nanotube is not fully covered by a monolayer of Li atoms (at the ratio Li:C=1:1), it is still possible to achieve high $H_2$ uptake as long as the Li:C ratio is higher than 1:2. Note also that the BC nanotube system studied in this work has a high B:C ratio of 1. In reality, nanobubes with smaller B:C ratio such as $B_xC$ (x≤0.10) and $BC_3$ nanotubes have been made [33-38]. Based on our model study, we expect that the binding energy of Li atom to the $BC_3$ nanotube should be also much larger than the cohesive energy of bulk lithium. Therefore, it is plausible to produce Li-dispersed $BC_3$ nanotubes, such as the $BC_3Li_2$ system. The latter has a hydrogen gravimetric density of 6.2 wt%, if one $H_2$ is adsorbed on every Li atom.

Besides the gravimetric density of hydrogen, the DOE target of volumetric density (> 45 g/L) should be also met. To this end, the boron carbide nanotubes must be modestly separated rather than in the form of bundles. Creating a pillared nanotube array is a possible solution as suggested by Deng *et al.* [17] [see Fig. 4(a)]. Lithium vapor can be injected to the pillar to coat the BC tube surface. Ideally, if the inter-tube distance is 16 Å [Fig. 4(b)], the nearest $H_2$-$H_2$ distance will be ~2.68 Å and the volumetric density of hydrogen will be ~50.5 g/L. If the inter-tube distance increases to 17 Å [Fig. 4(c)], the volumetric density of hydrogen will be 44.8 g/L, still close to the 2010 DOE target (45 g/L).

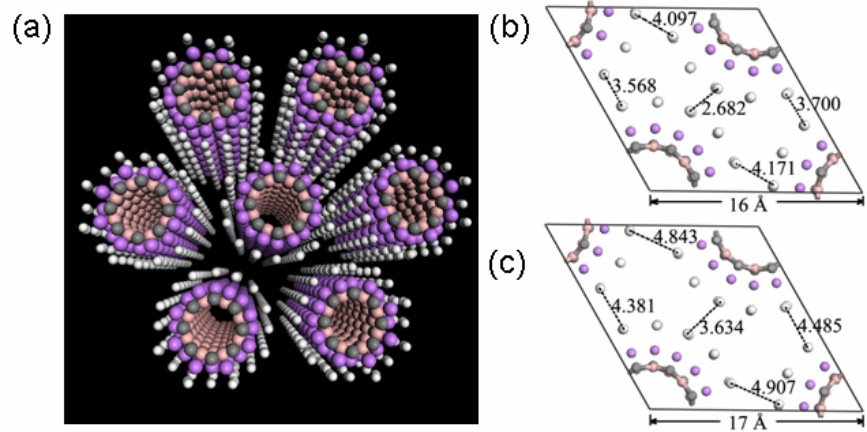

**FIG. 4 Hydrogen molecules are adsorbed on the outer surface of the pillared Li-dispersed (8,0) boron carbide nanotubes. (a) Hydrogen molecules are adsorbed in the Li-dispersed (8,0) BC nanotube pillar. The distance between the centers of two Li-dispersed BC nanotubes is (b) 16 Å and (c) 17 Å.**

**Conclusions**

We report results of density-functional calculations of hydrogen adsorption on Li-dispersed boron carbide nanotube. We have shown that the boron substitution greatly enhances the binding energy of Li atom to the nanotube, and the binding energy (~ 2.7 eV) is much greater than the cohesive energy of lithium metal (~1.7 eV). The much enhanced binding energy can assure higher lithium doping concentratino and yet lessen the tendency of lithium aggregation or segregation during lithium vapor deposition. We have also shown that pillared Li-dispersed boron carbide nanotubes are capable of high hydrogen uptake. With one hydrogen molecule adsorbed on every Li atom, the ideal BCLi system or the more practical $BC_3Li_2$ system can achieve hydrogen mass density of 6.2 wt%. Moreover, the average adsorption energy of hydrogen is likely in between 0.1 eV (GGA calculation) and 0.24 eV per $H_2$ (LDA calculation), close to 0.15 eV which is the optimal adsorption energy suggested for reversible adsorptive hydrogen storage at room temperature [27]. Finally, at inter-tube distance of 16 Å, the pillared Li-dispersed boron carbide nanotubes can achieve a volumetric density of hydrogen uptake ~50.5 g/L.

**Acknowledgments**

We are grateful to valuable discussions with Dr. S.B. Zhang and Professor J.L. Yang. This work is supported by grants from the DOE (DE-FG02-04ER46164), the Nebraska Research Initiative, and NSFC (#20628304), and by the Research Computing Facility at University of Nebraska-Lincoln.